# NEGATIVE DIFFERENTIAL RESISTANCE OF ELECTRONS IN GRAPHENE BARRIER

D. Dragoman – Univ. Bucharest, Physics Dept., P.O. Box MG-11, 077125 Bucharest, Romania

M. Dragoman[*] – National Institute for Research and Development in Microtechnology (IMT), P.O. Box 38-160, 023573 Bucharest, Romania

**Abstract**

The graphene is a native two-dimensional crystal material consisting of a single sheet of carbon atoms. In this unique one-atom-thick material, the electron transport is ballistic and is described by a quantum relativistic-like Dirac equation rather than by the Schrödinger equation. As a result, a graphene barrier behaves very differently compared to a common semiconductor barrier. We show that a single graphene barrier acts as a switch with a very high on-off ratio and displays a significant differential negative resistance, which promotes graphene as a key material in nanoelectronics.

_________________________________________________________________

a) Corresponding author. Email: mircead@imt.ro, mdragoman@yaoo.com

The graphene is a monolayer-thick sheet of graphite formed by a repetitive honeycomb lattice in which carbon atoms bond covalently with their neighbours. The graphene constitutes the basic structure of many carbon-based materials. For example, when the graphene is stacked it forms the graphite, and when it is rolled up it forms the carbon nanotube, which is a key material for many nanoelectronic devices working in a wide spectral range, from few hundred of MHz to X rays.

Until recently, it was alleged that graphene as a monolayer cannot exist in a free state, but experiments have succeeded to isolate such a carbon monolayer using mechanical exfoliation[1] from pyrolytic graphite and to deposit it on a $SiO_2$ layer grown on a doped Si substrate. The entire structure behaves like a field effect transistor if metallic electrodes are added to the graphene. Micromechanical manipulation techniques are also used to fabricate graphene of various dimensions.[2] These methods are able to produce large graphene sheets, up to 100 μm in size. In graphene the electron and hole states are correlated through charge-conjugation symmetry, their transport being described by a Dirac-like equation, which is a relativistic-like quantum equation for particles with spin ½. This is an astonishing situation since in common semiconductors and the heterostructures based on them two unrelated Schrödinger equations describe the electron and hole transport, respectively. There are also other striking differences between physicals effects in graphene and common semiconductors, such as Landau levels and Hall conductivity. In this respect, Ref. 3 is a very recent review dealing with the physical properties of graphene.

The dispersion relation in graphene is linear for both electrons and holes: $E = \pm|\hbar\mathbf{k}|v_F$, where $v_F$ is the Fermi velocity and the positive sign corresponds to electrons, while the negative sign is assigned to holes. This unusual dispersion relation indicates that graphene is a gapless semiconductor because the conduction and valence bands touch in one point, often termed Dirac point. In addition, this linear dispersion relation in graphene is encountered in nature in only one other case: for photons. However, the linearity of the dispersion relation for electrons in graphene and photons, has very different physical significances. In the case of electrons transported through graphene, the linearity of the dispersion relation indicates that the effective mass of electron and holes is zero. The vanishing of the effective mass signifies that there is no interaction between electrons and holes and the lattice of the graphene, so that electrons propagate balistically through graphene with the velocity $v_F \cong c/300$, where $c$ is the speed of light. This is an important characteristic of graphene, which will be used further to analyze graphene-based quantum devices. In the case of photons, which are bosons, the linearity of the dispersion relation $E = \hbar\omega = hc/\lambda$ expresses photon propagation in vacuum with the speed of light $c$. From this perspective, the graphene is a slow-wave structure in which electrons and holes propagate with a velocity that is at least two orders of magnitude slower than $c$.

We will focus further on the tunnelling effect in graphene. It is well known that in semiconductor heterostructures the transmission through a barrier is strongly dependent on (in fact, decays exponentially with) the barrier width and height, because inside the barrier the wavenumber is purely imaginary and corresponds to evanescent propagation. The transmission can equal 1 only in multilayer structures, in which well layers are sandwiched between barriers, and only when the energy of electrons equals a resonant level of the quantum well. This effect is called resonant tunnelling and has very important applications in high-speed devices such as resonant tunnelling diodes and semiconductor cascade lasers. [4]

Due to the Dirac equation for massless fermions in graphene, the tunnelling through a barrier in graphene is described by the Klein tunnelling mechanism, also known as Klein paradox. In graphene, in deep contrast with semiconductor heterostructures, if the ballistic electrons are normally incident on the barrier, the transmission is 1 irrespective of the barrier height and width since electron propagation is not evanescent inside the barrier region.[5] More precisely, holes take the role of electrons as charge carriers in barriers. So, at least from the point of view of the transmission value, a single graphene barrier is equivalent to a more complicated resonant tunnelling device in heterostructures made from common semiconductors. The aim of this paper is to demonstrate that this equivalence can be generalized to other phenomena, and in particular to show that a graphene barrier displays a differential negative resistance, like any resonant tunnelling structure. The evaluation of the traversal time through graphene barrier devices reveals that graphene could play a major role in high-speed devices and in the generation of very high frequencies exceeding few THz.

The Dirac spinors $\Psi_1$ and $\Psi_2$ for an electron wave incident from region 1 on a barrier of height $V_0$ and width $D$ and propagating at an angle $\varphi_1$ with respect to the $x$ axis are given by

$$\Psi_1(x,y) = \begin{cases} [\exp(ik_1 x) + r\exp(-ik_1 x)]\exp(ik_y y), & x \leq 0 \\ [a\exp(ik_2 x) + b\exp(-ik_2 x)]\exp(ik_y y), & 0 < x < D \\ t\exp(ik_3 x)\exp(ik_y y), & x \geq D \end{cases} \tag{1a}$$

$$\Psi_2(x,y) = \begin{cases} s_1[\exp(ik_1 x + i\varphi_1) - r\exp(-ik_1 x - i\varphi_1)]\exp(ik_y y), & x \leq 0 \\ s_2[a\exp(ik_2 x + i\varphi_2) - b\exp(-ik_2 x - i\varphi_2)]\exp(ik_y y), & 0 < x < D \\ s_3 t\exp(ik_3 x + i\varphi_3)\exp(ik_y y), & x \geq D \end{cases} \tag{1b}$$

In these expressions, which generalize those in Ref. 5 for the case when a bias $V$ is applied over the structure, $k_y = k_F \sin\varphi_1$ with $k_F$ the Fermi wavenumber, $k_1 = k_F \cos\varphi_1$, $k_2 = [(E - V_0 + eV/2)^2 / \hbar^2 v_F^2 - k_y^2]^{1/2}$, $k_3 = [(E + eV)^2 / \hbar^2 v_F^2 - k_y^2]^{1/2}$, $\varphi_{2,3} = \tan^{-1}(k_y / k_{2,3})$, $s_1 = \text{sgn}\, E$, $s_2 = \text{sgn}(E - V_0 + eV/2)$, and $s_3 = \text{sgn}(E + eV)$, with $E$ the electron energy. The transmission coefficient through the barrier $T = s_3 \cos(\varphi_3) |t|^2 / s_1 \cos(\varphi_1)$ is determined by imposing the requirement of wavefunction continuity at the $x = 0$ and $x = D$ interfaces. The

barrier can be created by p-doping or, simpler, by gating.[5] The potential energy diagram of the biased barrier surrounded by non-gated regions is illustrated in Fig.1. For ease of computation we have approximated the linear potential drop across the barrier with a step-like drop of average value $eV/2$.

Before illustrating the behaviour of the transmission coefficient with several parameters, we must emphasize once again the peculiar charge transport mechanism in graphene. In the regions where the electron energy $E$ is higher than the potential energy, as is the case in the first and last region in Fig. 1, charge transport is assured by electrons, whereas in the barrier region, when the electron energy is lower than the potential energy, holes assume the charge transport role. The term "barrier" does not indicate in this case a region of evanescent propagation, as in common semiconductors, but a region in which charge transport is undertaken by holes instead of electrons. Therefore, although it is mathematically possible that for certain combinations of the parameters $V_0$, $V$, $v_F$ and $\varphi_1$ the wavevector $k_2$ becomes imaginary, this situation does not indicate evanescent propagation, which is forbidden, but total reflection of the electron wavefunction at the barrier boundary: a graphene barrier, irrespective of its thickness, can act in certain situations as a classical barrier for a quantum wavefunction in the sense that the wavefunction cannot penetrate (not even exponentially) inside the barrier. This peculiarity of charge transport in graphene cannot be emphasized enough.

In Fig. 2 the transmission of the graphene barrier is displayed as a function of voltage at various Fermi wavenumbers $k_F = \alpha k_{F0}$, where $\alpha$ is 0.25, 0.3 and 0.35 when $\varphi_1 = 15^0$ and $D = 100$ nm. The parameter $k_{F0}$ corresponds to a Fermi wavelength of 50 nm. The most striking feature of these curves is the fact that the transmission has a gap, which increases when $k_F$ is increasing. This happens because the transmission of the incident wavefunction is forbidden in the barrier because otherwise $k_2$ would be imaginary, fact that is not allowed by the gapless band energy diagram of the graphene. This means that graphene is a native quantum switch of ballistic electrons. Such an abrupt ON-OFF characteristic is unprecedented in nanoelectronic devices. The behaviour of the transmission coefficient with $D$ and the angle of incidence is similar to that in Ref. 5, so that we do not reproduce these curves here.

Further, in Fig. 3 the $I-V$ characteristics of the graphene barrier at room temperature are displayed for the structures considered in Fig. 2: curves are plotted at various Fermi wavenumbers $k_F$, when $\varphi_1 = 15^0$ and $D = 100$ nm. The $I-V$ curve was calculated using the Landauer formula. We can see from Fig. 3 that a very pronounced negative differential resistance region is occurring. Increasing $k_F$, the peak-to-valley ratio of this region is increasing in the range 2.5–4 in our simulations, but the maximum value of current is decreasing. Different Fermi wavenumbers can be selected by n-doping.

In Fig. 4, for $k_F = 0.3\, k_{F0}$, we have represented the $I-V$ dependence at various incident angles for $D = 100$ nm and at room temperature. We can see again that increasing $\varphi_1$, the peak-to-valley ratio of this region is increasing in the range 2.5–4, but the maximum value of the current is decreasing. Different incidence angles can be achieved by tilting the gate that delineates the barrier region with respect to the electron propagation direction, as suggested in Ref. 5, or by employing tilted electron emitters.[6]

The existence of the negative differential resistance region is due to the gap in the transmission coefficient. This is exactly the opposite situation to resonant tunnelling diodes made of semiconductor heterostructures, where the physical origin of the peak in the current is due to a narrow peak in transmission.

The negative differential resistance is the physical mechanism at the heart of generation of monochromatic signals of high frequencies, up to THz, as described in Ref.4. A device that generates such frequencies, called oscillator, consists of the negative differential resistance device loaded with a load, which selects the desired oscillation frequency. Taking into account that the transport in graphene is ballistic, the cut-off frequency of such a device is $f_c = v_F / 2\pi D \cong c / 600\pi D$, which for $D = 100$ nm gives 1.6 THz. This cut-off frequency could be increased decreasing the barrier width. Simulations show that a decrease in the barrier width does not significantly affect the $I - V$ characteristics.

In conclusion, the two-dimensional graphene barrier in which electrons are transported ballistically and are described as massless fermions (solutions of the Dirac equation) shows very abrupt ON-OFF transmission characteristics for appropriate values of its parameters and displays a significant negative differential resistance region. Both these characteristics are not encountered in any single layer of semiconductor material in which electron transport is described by the Schrödinger equation. Therefore, the transport properties of graphene are unique and single barrier layers of this material can be used instead of multiplayer common semiconductor devices for more compact integrated devices. In particular, unexpected applications for nanoscale high frequency devices, can be envisaged for graphene barriers.

**Figure captions**

Fig. 1 Energy band diagram for a biased graphene barrier.

Fig. 2 Transmission of the graphene barrier as a function of the applied voltage at various Fermi wavenumbers: $k_F = 0.25\ k_{F0}$ (dotted line), $k_F = 0.3\ k_{F0}$ (solid line), and $k_F = 0.35\ k_{F0}$ (dashed line).

Fig. 3 The *I-V* characteristics of graphene barrier at various Fermi wavenumbers: $k_F = 0.25\ k_{F0}$ (dotted line), $k_F = 0.3\ k_{F0}$ (solid line), and $k_F = 0.35\ k_{F0}$ (dashed line).

Fig. 4 The *I-V* characteristics of graphene barrier at various incident angles: $\varphi_1 = 10^0$ (dotted line), $\varphi_1 = 15^0$ (solid line), and $\varphi_1 = 20^0$ (dashed line).

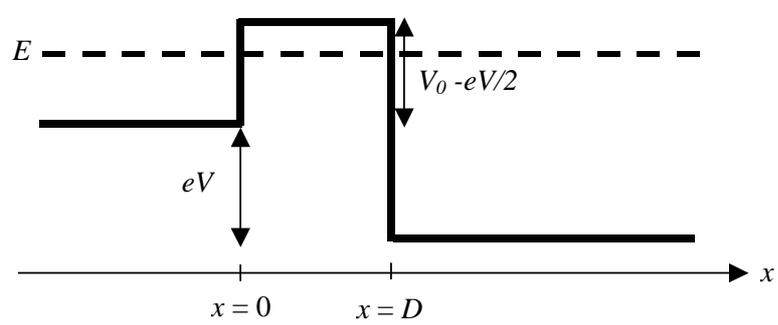

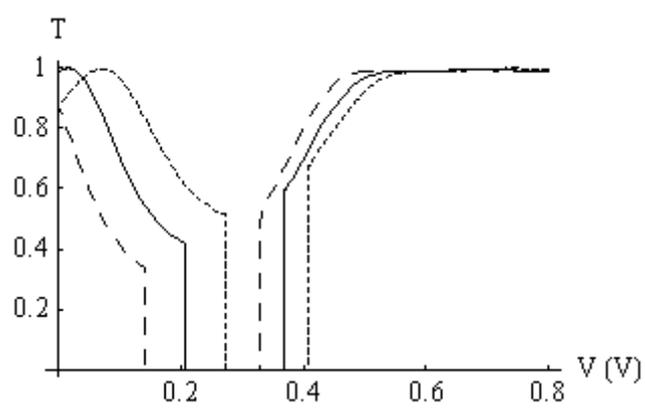

Fig.2

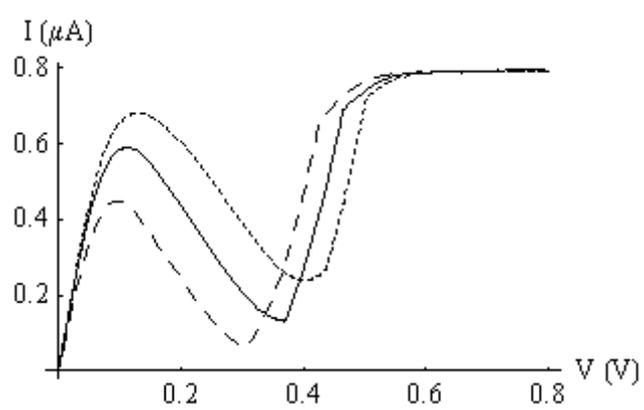

Fig.3

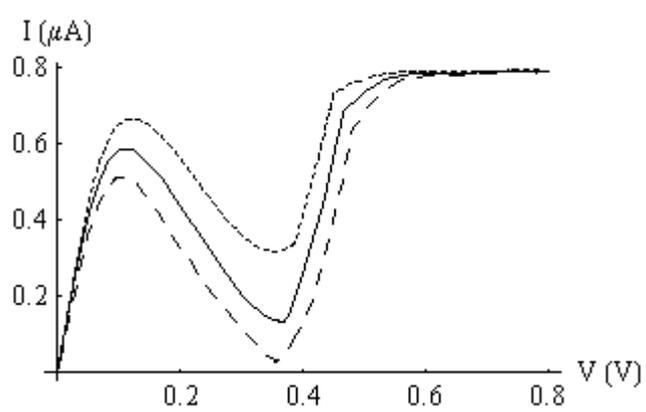

Fig.4